# Emerging ML-AI Techniques for Analog and RF EDA

Zhengfeng Wu
Drexel University
Philadelphia, Pennsylvania, USA
jw3723@drexel.com

Nnaemeka Achebe
Drexel University
Philadelphia, Pennsylvania, USA
nma334@drexel.edu

Ziyi Chen
Drexel University
Philadelphia, Pennsylvania, USA
zc372@drexel.edu

Vaibhav V. Rao
Drexel University
Philadelphia, Pennsylvania, USA
vv85@drexel.edu

Pratik Shrestha
Drexel University
Philadelphia, Pennsylvania, USA
ps937@drexel.com

Ioannis Savidis
Drexel University
Philadelphia, Pennsylvania, USA
is338@drexel.com

**Abstract**

This survey explores the integration of machine learning (ML) into EDA workflows for analog and RF circuits, addressing challenges unique to analog design, which include complex constraints, non-linear design spaces, and high computational costs. State-of-the-art learning and optimization techniques are reviewed for circuit tasks such as constraint formulation, topology generation, device modeling, sizing, placement, and routing. The survey highlights the capability of ML to enhance automation, improve design quality, and reduce time-to-market while meeting the target specifications of an analog or RF circuit. Emerging trends and cross-cutting challenges, including robustness to variations and considerations of interconnect parasitics, are also discussed.

## 1 Introduction to ML-AI Techniques for Analog/RF Design Automation

Analog design remains a cornerstone of modern integrated circuits, accounting for approximately 20% of the chip area and 40% of the total IC design effort [1]. In addition, analog circuits contribute to approximately 50% of the costly design iterations that occur during development [1]. As analog and RF systems evolve toward higher frequencies and greater levels of integration, traditional knowledge-driven methods struggle to address the increasing computational and design complexities [2].

Electronic design automation (EDA) has transformed the design of an IC that allow for high-level circuit implementation strategies. However, while digital design has achieved high levels of abstraction and automation, analog design automation lags behind as analog circuits are highly customized [3]. Analog synthesis and physical design typically follows a hierarchical flow that includes topology generation, device sizing, layout generation, and post-layout verification. The challenges of automating analog design stem from the highly non-linear design space, computational complexity, and stringent performance and manufacturing constraints, which result in complex multi-objective optimization problems that require intricate trade-offs between competing circuit objectives [3].

Recent advances in machine learning (ML) offer promising solutions, allowing for the accurate prediction of critical circuit parameters and the guidance of early-stage design decisions. Data-driven methods have been proposed to enhance productivity and improve the quality of design provided by EDA workflows.

This survey explores the integration of ML in analog EDA, covering both synthesis and physical design tasks. In Section II, a review of learning and optimization algorithms is provided. A discussion of ML-based algorithms for analog circuit synthesis and physical design tasks is provided in Section III. Cross-cutting challenges, including parasitic-aware design and variation effects are described in Section IV. Finally, the importance of standardized benchmarks and datasets to drive further innovation in ML-driven analog EDA is discussed in Section V.

## 2 Overview of Learning and Optimization Algorithms for Analog EDA

In this section, an overview is provided on the learning and optimization algorithms utilized for analog circuit design. Learning models map from circuit features and design variables to target performance parameters, enabling informed decisions during early design stages and reducing reliance on costly simulations. The learning models are usually applied as surrogates of the design space that guide optimization algorithms.

Learning techniques including statistical models, neural networks, and transfer learning for circuit modeling tasks are introduced in Section II-A. In Section II-B, methods including Bayesian optimization and reinforcement learning are described that are utilized to optimize complex analog design spaces. An overview of applying ML algorithms for the synthesis and physical design of analog circuits is shown in Fig. 1.

### 2.1 Learning

Learning algorithms include statistical and neural network based methods. A description of both follows.

*2.1.1 Statistical Learning Algorithms* Statistical learning methods, including linear regression, support vector machines [4], and tree-based models such as random forests and gradient boosting [5], have been widely used for analog design tasks. The statistical learning methods have proven to be effective when computational resources are limited and available data is sparse. The interpretability provided by statistical learning models, including the analysis of the importance of features and the visualization of decision processes such as the structure of a tree, provide benefit when analyzing the relationships amongst design parameters [5, 6].

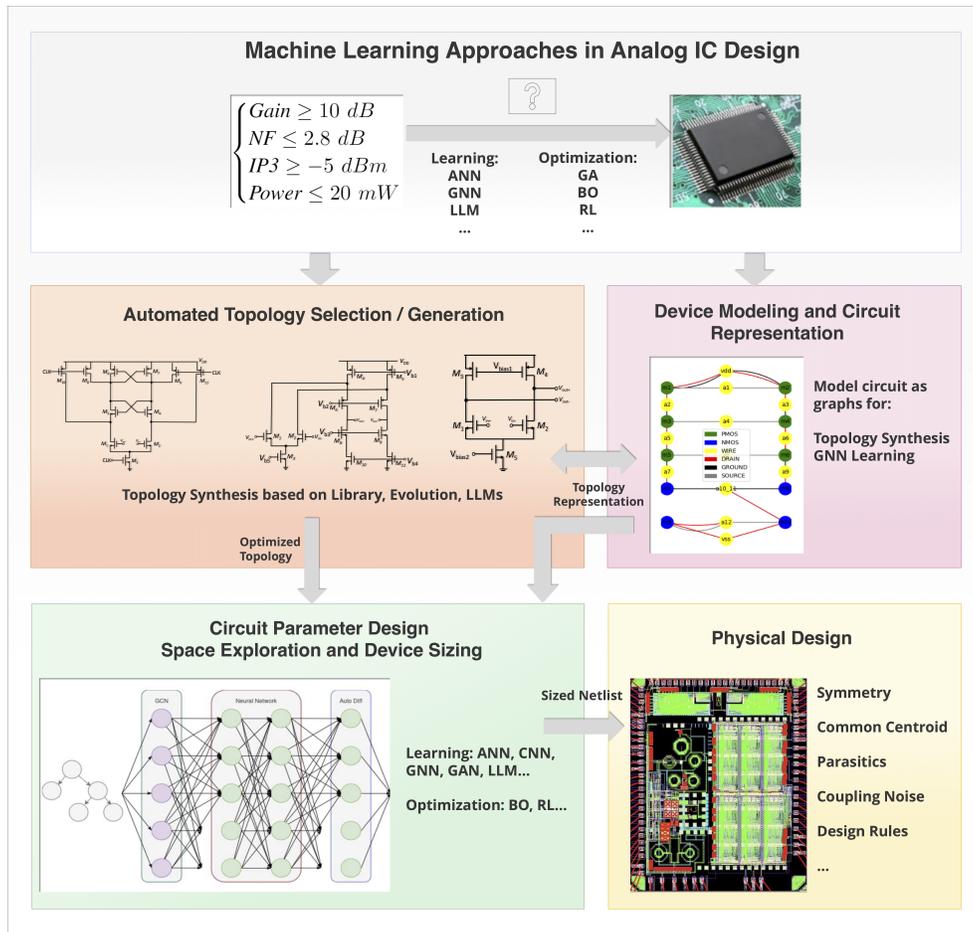

Figure 1: An overview of applying machine learning for the synthesis and physical design of an analog and RF circuit.

*2.1.2 Neural Network-Based Approaches* Over the past ten years, neural network-based approaches have gained traction in analog design automation, beginning with the versatile multi-layer perceptron (MLP) [7]. More recently, graph neural networks (GNNs) [8–11] have shown significant promise in learning circuit connectivity and topological features. For graph representations to model analog circuit behaviors at the device level, vertices often represent active and passive devices, while edges represent nets [8, 9]. However, some models treat both devices and nets as nodes when interconnect features are emphasized for better prediction or optimization performance [10, 11]. Generative adversarial networks (GAN) have also been applied to guide the routing of an analog IC [12].

Transfer learning permits the reuse of pre-trained model layers on new tasks [13], during either standalone model training [14] or execution of optimization [15]. Transfer learning allows for the generalization of models across circuit design objectives including for different technology nodes and different circuit topologies [14–16], which reduces data requirements and improves model performance.

## 2.2 Optimization

Two classes of optimization algorithms include classic methods and surrogate-assisted. A discussion of each follows.

*2.2.1 Classic Optimization Algorithms* Algorithms based on gradient descent are computationally efficient and are often used for problems with differentiable cost functions [17]. However, the algorithms are limited by potential convergence to local minima, especially in non-convex design spaces.

Optimization utilizing heuristic-based evolution algorithms, including simulated annealing, particle swarm optimization, and genetic algorithms [18], allow for a global exploration of the design space. Genetic algorithms apply principles of evolution to iteratively refine populations of design candidates through crossover and mutation operations [18].

*2.2.2 Surrogate-Assisted Optimization* Algorithms that perform surrogate-assisted optimization, including Bayesian optimization (BO) [19, 20], more efficiently explore the design space by utilizing surrogate models like Gaussian processes to approximate expensive cost functions. The search process, therefore, proceeds with fewer costly evaluations of the design space.

Reinforcement learning (RL) [15, 21] has been explored for optimization in a sequentially executed decision-making flow. In RL, an agent iteratively interacts with the circuit simulation environment and refines actions to maximize cumulative rewards based on results returned from the simulation framework.

## 2.3 Large Language Models for Analog EDA

Since 2023, large language models (LLMs) have emerged as promising tools to address the challenges in design [22–24]. LLMs excel in processing unstructured inputs, such as textual specifications, and translating high-level requirements into actionable design strategies.

LLMs have been applied to recognize patterns in circuit data, predict relationships between parameters and performance, generate topologies, and derive sizing specifications based on prior knowledge [22–24]. In addition, LLMs integrate with optimization algorithms including BO and RL to iteratively determine and refine solutions that meet design objectives [22, 23]. LLMs also provide an intuitive and accessible interface, lowering the barrier to entry for EDA tool users.

## 3 Data-Driven and Heuristic Approaches for Analog Circuit Synthesis and Physical Design

Analog and RF circuit synthesis and physical design involve a diverse set of tasks, which include defining design constraints, generating topologies, modeling devices, sizing circuits, and optimizing layouts. Unlike digital design, nuanced constraints such as device matching, symmetry, parasitic effects, and noise isolation must be considered. An overview of heuristic and ML-based approaches for analog/RF synthesis and physical design is provided in this section.

### 3.1 Defining Constraints and Design Specifications

Primitives including differential pairs and current mirrors form the basis of an analog circuit. Accurate recognition of circuit hierarchies is required to properly extract constraints on the symmetry of devices when generating a physical layout. The detected symmetrical device groupings also provide supplemental information when utilized as features in ML models for downstream circuit modeling and optimization tasks.

Traditional methods rely on isomorphism matching between circuit graphs and primitive libraries [25, 26], with some approaches dynamically constructing libraries during execution [27]. More recently, learning-based methods that utilize GNNs have been proposed to classify hierarchical circuit levels [10], predict graph edit distances [28], and detect circuit substructures [29]. Heterogeneous GNNs predict device and system-level symmetries by capturing topological features [30]. To return the specific circuit category of a detected functional group, hybrid approaches that combine GNNs with subgraph isomorphism[31] have been proposed. A comparative analysis of heuristic isomorphism-based and learning-based algorithms that recognize hierarchy is provided in [32].

### 3.2 Automated Topology Generation for Analog Circuits

Given a set of specifications, the design of an analog circuit begins with the synthesis of a circuit topology, which traditionally requires extensive human expertise and intuition [33]. Automating topology selection and generation is computationally expensive due to the large search space [34]. Exploring all possible topologies is neither practical nor efficient. Although the infusion of prior knowledge through a pre-defined topology library narrows the design space, the discovery of novel topologies is limited. A scalable and efficient solution for topology synthesis in analog design is yet to be provided.

Techniques that generate analog circuit topologies are categorized into knowledge-based methods and evolutionary algorithm-based methods [34]. With knowledge-based methods, circuit components are systematically assembled according to predefined rules. Bell Laboratories Analog Design Expert System (BLADES) [35] is a knowledge-based algorithm that integrates formal mathematical methods with heuristic reasoning. A graph-based circuit topology generator is proposed that generates candidate topologies as hierarchical tree structures, guided by graph grammar rules and building blocks from a predefined library [36].

Evolutionary algorithms utilize a stochastic exploration process to address high-dimensional, discrete multi-objective tasks. In [37], circuit topologies are encoded as connection matrices and value vectors, where crossover and mutation are applied to explore a vast design space of up to $10^{19}$ possible configurations. In [33], a segmented evolution strategy is proposed that progressively refines the topology using a genetic algorithm while optimizing for objectives that include performance, component reduction, and area.

Hybrid approaches that integrate both heuristic and ML methods are increasingly being explored. A technique is proposed [21] that combines reinforcement learning techniques with a policy gradient neural network to iteratively construct circuit topologies using predefined building blocks, while the performance is evaluated using symbolic analysis and SPICE simulations.

### 3.3 Device Modeling and Synthesis of EM Structures

Modeling, design, and optimization of active and passive devices as well as electromagnetic (EM) structures is required for analog/RF circuit design. Traditional modeling of EM structures relies on highly accurate mathematical calculations, which include use of finite element methods (FEM) and method of moments (MoM) to compute EM fields. However, the solvers are computationally costly, particularly for novel devices such as FinFET and gate-all-around transistors, which hinders design technology co-optimization (DTCO)[38]. ML has emerged as an effective tool to enhance traditional modeling methods by improving prediction accuracy and reducing computational cost.

ML has been applied for the accurate modeling and representation of parameters of devices and EM structures, where parameters including resonant frequency, bandwidth, and impedance directly impact performance. In [39], ML is applied to predict the current and capacitance of FinFETs based on other device parameters. An autoencoder-based approach is utilized to develop a PIN diode model, utilizing unsupervised learning to compress high-dimensional data into a latent space[40], which accelerates the extraction of device parameters. In [41], an ANN is introduced that performs real-time BSIM parameter extraction in nanosheet FETs, while accounting for multiple structural variations.

For the modeling of EM structures, an on-chip transformer automatic synthesis (OTAS) flow is proposed in [42] that utilizes Gaussian process regression models to automate the translation of system requirements into transformer design parameters, reducing the effort required for impedance matching. Similarly, recent work on deep learning-enabled mmWave power amplifier (PA) and antenna design has shown translations from high-level design specifications to physical layouts[43].

Physics-based device equations have been integrated with ML models to align model predictions with known device behaviors. In [44], analytical equations process variations such as current shifts and threshold voltage shifts are integrated into the learning models. A graph-based compact model (GCM) is proposed in [45] that represents physical parameters, including threshold voltages and channel length dependence, as graph nodes. Implemented in Verilog-A, GCM provides accurate predictions with 300 sample points, while passing industry-standard benchmark tests. GCM is integrated with SPICE simulations, offering a computationally efficient approach to advanced transistor modeling.

## 3.4 Analog/RF Performance Modeling and Device Sizing

Determining the optimal sizes of analog devices, including transistors, diodes, resistors, capacitors, and inductors, ensures that the chosen topology satisfies the design specifications. The early methods that sized devices are based on solving analytical equations that establish relationships between design parameters and performance metrics, where multi-objective optimization problems are formulated [46] [18]. As technology scales, knowledge-based methods face limitations due to the growing complexity of circuit equations and the effort required to reconcile discrepancies between theoretical optimizations and simulation results.

Using data from simulation, data-driven techniques are utilized in a bottom-up approach to model circuit behaviors and extract design insights. Early automated sizing techniques are based on statistical learning methods [2, 47]. In [47], a statistical learning method is proposed that combines Kriging metamodeling with simulated annealing to optimize a sense amplifier. In CALT [48], random forest classifiers are used as surrogate models of circuit parameters and genetic algorithms are used to predict whether designs meet specified performance thresholds. More recently, neural networks have been applied with optimization algorithms [7, 49].

Reinforcement learning [15, 50, 51] and Bayesian optimization [20] are state-of-the-art optimization engines for analog device sizing. With the RL framework proposed in DNN-Opt [51], an actor network is utilized for tuning design parameters and a critic network is applied for predicting performance parameters, which results in up to a 24x reduction in the number of executed SPICE simulations. GNNs are integrated into RL-based search in [15] and [50], where circuit topology information is included for optimization.

To accelerate the execution of sizing algorithms, techniques have been proposed that allow parallelization. When the simulation time for a given set of design variables is constant, parallel execution is leveraged for algorithms compatible with multi-core simulators, whereas those relying on sequential decision-making remain serial. A batch-constrained Bayesian optimization (BO) methodology is proposed in [20] that utilizes a multi-objective acquisition function ensemble as a substitute of the sequential execution of traditional BO. An asynchronously parallel batch optimization method for analog sizing is proposed in [52] that utilizes a ranking approximation method to select between cheap and expensive simulations of circuit parameters.

Another emerging research direction is layout-driven device sizing [52], which includes an iterative design loop between schematic-level sizing and layout generation, minimizing the need for additional post-layout adjustments.

## 3.5 Placement Optimization in Analog Design

Analog IC placement involves determining the optimal locations for devices within a given circuit topology to optimize specific performance metrics. The placement algorithm must adhere to topological constraints, including device matching, symmetry between devices or device groups, and proximity requirements. Depending on the target application, constraints that include alignment, thermal gradients, or current/signal flows must also be met [3].

Traditional analog circuit placement methodologies have been mostly manual, leading to an increase in circuit development time and cost. ML-based approaches proposed in [53–56] utilize ANNs and deep generative models to provide an optimum placement of a circuit while considering topological constraints. In [53, 54], a nonlinear ANN model is proposed that is trained with semi-supervised learning. The ANN outputs the placement coordinates of each cell (sub-block) for a given sizing and set of topological constraints. DeepPlacer, a deep learning generative model, is proposed that performs the placement of multiple amplifier topologies in less than 150 ms across different technology nodes [55], proving the scalability of ML-based placement methodologies.

Unsupervised models including AGraph2Seq introduce structure through an attention-based graph-to-sequence encoder-decoder architecture that enables efficient placement of analog circuit blocks with a minimal set of trainable parameters [56]. The AGraph2Seq methodology encodes the topological constraints of the circuit using relational GCNs and a long short-term memory (LSTM) decoder, which outputs the position of a device considering the relative positions of the devices that have already been placed.

Graph-based ML techniques proposed in [57] and [58] explore hierarchical circuit decomposition and identifying symmetry constraints for physical placement. In [57], a GNN based on a neural tensor network is integrated with exact graph matching to account for multi-level symmetrical hierarchies when performing placement. In [12], *WellGAN* is proposed, which is a generative adversarial network (GAN) that is trained on high quality layouts that are manually generated. A conditional GAN is utilized to generate analog layouts that mimick the target circuit behavior.

For RF placement, a depth-first device placement strategy is proposed in [59] that utilizes parameterized cells. A decision tree model identifies the coordinates of placed devices, rotation angles, and mirror operations. Precise pin positions and alignment in routing are computed using analytical models. Constraints including impedance matching and electromagnetic performance are considered. For instance, a 5-ring LNA (6–13 GHz) was placed in 86 ms. The generated RF layout provides improved performance over layouts generated with Keysight ADS.

Millimeter-wave power amplifiers, passives, and EM structures are physically placed utilizing a deep-learning-based inverse design framework [60]. A convolutional neural network (CNN) is utilized to predict scattering parameters for 2 D planar structures, while a genetic algorithm optimizes the spatial configuration [60]. The generated power amplifier achieves a power added efficiency (PAE) of 16%-24.7% across a 3-dB bandwidth of 30-94 GHz. The approach expands the design space beyond pre-defined templates, achieving efficiency performance in less time.

## 3.6 Routing Strategies in Analog/RF Design Automation

Routing significantly impacts the performance and manufacturability of a circuit. While the core objective of routing algorithms is to find the shortest physical path between two connected nodes of a graph, routing for analog and RF circuits must address additional performance-critical constraints including symmetry, minimization of parasitic impedances, and noise isolation, while adhering to geometric and electrical design rules. Analog routing is, therefore, a complex optimization problem despite the smaller number of devices. In addition, for fabrication technology nodes below 22 nm, the complexity of design rules has increased nearly tenfold, surpassing 10,000 rules [61]. The increase in circuit and design constraints requires novel methodologies to efficiently achieve optimal routing of analog and RF circuits.

Analog routing techniques share similarities with digital routing, utilizing maze routing, A*, integer linear programming (ILP), SAT-based, and ML-guided algorithms. Common-centroid (CC) routing [62] is introduced to mitigate mismatch due to systematic variations. Circuit layouts that utilize CC routing ensure that systematic variations across a linear gradient are mitigated by symmetrically placed elements. However, CC layouts often introduce complex routing challenges and increased parasitic effects due to the intricate routing paths required to interconnect symmetric elements.

BAG2 [46] integrates modularity and process portability into the routing strategy, using separate engines to automate wire placement, spacing, and layer selection. By isolating process-specific details with the developed parametric workflow, BAG2 adapts designs across technology nodes while balancing automated layout generation with designer input. Similarly, LAYGO [63] utilizes a grid-based and template-driven routing algorithm developed for sub-28 nm technology nodes. The algorithm abstracts the complexity of design rules by using predefined templates, grids, and relative placements to achieve design rule compliance and improved routing results.

A two-step approach is implemented in ALG [64], which combines global and local routing with sensitivity analysis. Global routing ensures adherence to performance constraints including symmetry and matching, while local maze routing optimizes paths while accounting for complex constraints that include crosstalk and planarity. While effective, maze routing sacrifices efficiency for precision, which results in potential trade-offs in wirelength and algorithmic execution time.

MAGICAL [65] integrates constraint-aware and symmetry-driven routing with placement. Using an A* algorithm for detailed routing, symmetry constraints extracted from the netlist are enforced, ensuring electrical balance in critical nets such as for differential pairs. Based on the base router, GeniusRoute [66] leverages variational autoencoders to mimic human design expertise, combining probabilistic routing path selection with the detailed routing provided by A*. The hybrid approach is robust and scalable, however, there is a dependence on the quality of the training data and heuristic functions.

ALIGN [3] utilizes a hierarchical routing method that combines ILP and SAT algorithms. GNNs are utilized to extract and enforce electrical and geometric constraints [3]. As a result, a global optimization of symmetry, coupling noise, and length matching is provided. However, the hierarchical routing algorithm is less scalable due to the computational expense of solving large optimization problems.

In [67], a variational autoencoder (VAE) is utilized to learn and generate routes for analog building blocks. The routing paths are extracted from legacy layouts. The VAE models are trained to encode placement and routing features.

By applying LLMs, GLayout [24] translates human language input of analog layout specifications into compact technology-generic layouts. Using retrieval augmented generation (RAG) and fine-tuned LLMs, GLayout achieves 70% task completion with 44% of generated layouts passing DRC and LVS verification for circuits containing blocks with up to 4 transistors.

## 4 Cross-Cutting Challenges in Analog/RF Design

Analog and RF circuit design faces cross-cutting challenges that impact all stages of the design flow. Key challenges include variations in process, voltage, and temperature (PVT), as well as variations due to aging, parasitics, and manufacturing effects. A survey of techniques that address the cross-cutting challenges is provided.

### 4.1 Designing for Variation and Yield

A robust analog circuit compensates for process, voltage, temperature, and aging variations to provide consistent performance that matches the target specifications. RobustAnalog[68] utilizes a multi-tasking RL engine, where each type of variation is assigned as a single task. A clustering algorithm is executed that groups tasks based on the difference between the current performance of the circuit and the target performance.

In [69], the authors propose a simulation-based optimization framework for analog circuit sizing to ensure that both the target performance and the robustness specifications are met by limiting variations in performance across process and temperature corners. In [70], transistor-level simulations, geometric programming, and projection-based performance modeling are utilized to optimize circuits when considering process and environmental variations, which results in a trade-off between yield and performance. In [71], adaptive response surface modeling and structural homotopy are utilized for globally reliable, variation-aware sizing, leveraging stochastic gradient boosting and automated execution of SPICE simulations to optimize a complex, high-dimensional design space.

### 4.2 Modeling of Parasitic Impedance in Analog/RF Circuits

The parasitic extraction (PEX) of a circuit layout provides accuracy at the cost of high computational overhead [73]. The capacitance of an interconnect is reported in standard formats that include DSPF and SPEF. A simplified model of the total capacitance of an interconnect consists of a lumped value to ground and the coupling between neighboring lines.

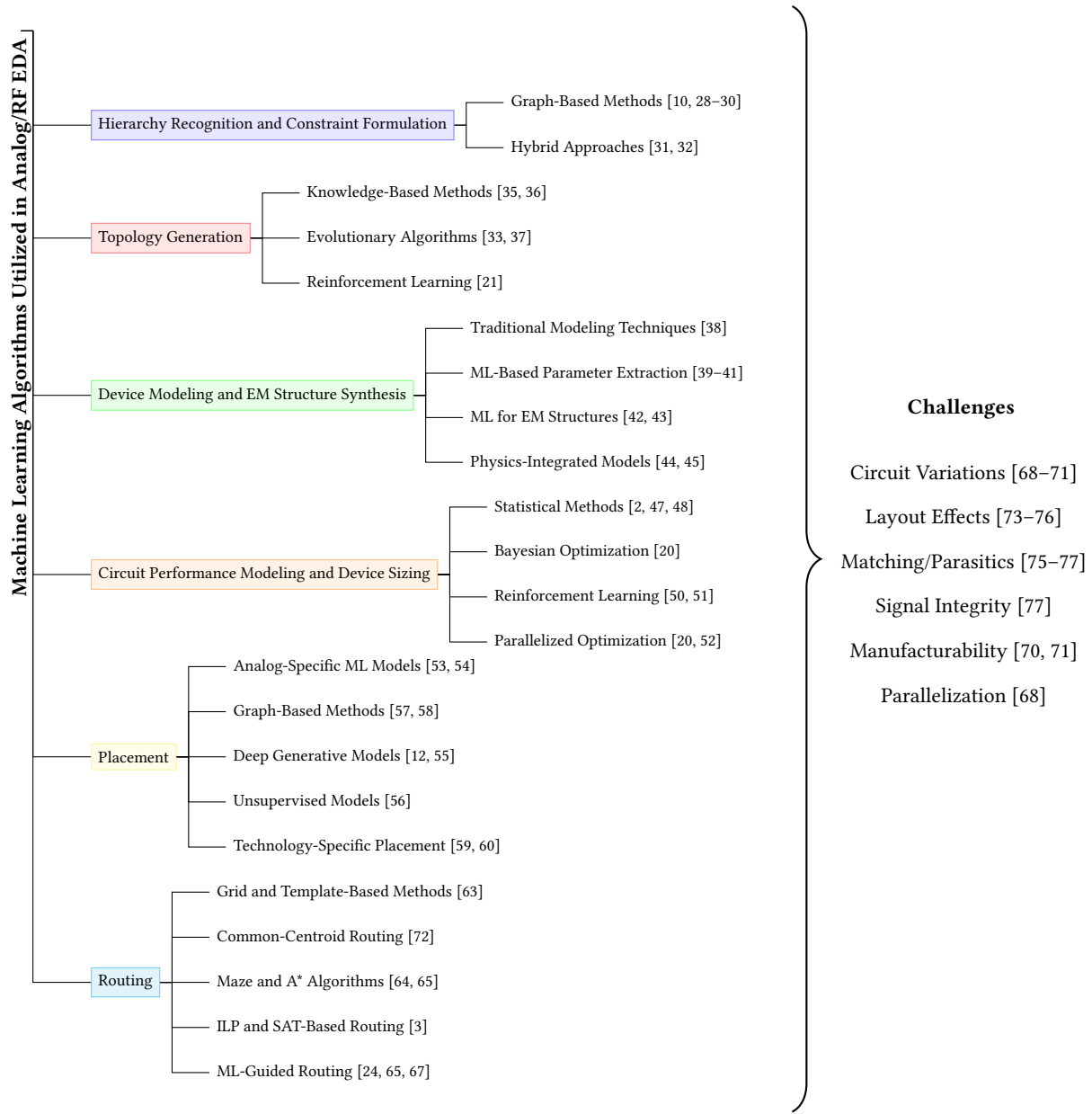

Figure 2: Overview of ML applications and algorithms for analog/RF circuit design. Cross-cutting challenges that impact all tasks are also highlighted.

ML has been applied to predict interconnect impedance in digital circuits [78, 79]. For analog design, random forests [74] and GNNs [75] have been applied for schematic-level parasitic estimation, capturing both device and interconnect parameters. Recent papers incorporate spatial features within the GNN module for post-placement capacitance prediction, utilizing self-attention mechanisms to predict the pairwise distances between devices [76] and edge-weighted GNNs learning from the pairwise distances between devices [75]. The ML-based methods improve efficiency by reducing dependence on PEX, while providing accurate estimates of parasitics at multiple design stages and combining schematic and spatial features. An accurate prediction of the coupling capacitance is significantly more challenging than estimating the physical capacitance to ground. A classifier is utilized in [77] to identify heavily coupled nets prior to routing, which allows for the early mitigation of flagged timing paths.

While ML-based interconnect models are effective for low-frequency (a few GHz) analog circuits, the models are inadequate at RF frequencies as interconnects behave like transmission lines rather than lumped resistors or capacitors. Dedicated matching networks, such as distributed *LC* component models and transmission line stubs, are implemented to achieve proper impedance matching, while ensuring the maximum power transfer and the minimal signal reflection [80]. Surrogate models of Gaussian processes are utilized to represent a high-speed channel, a millimeter-wave filter, and a low-noise amplifier [80]. Despite advancements in ML-based parasitic modeling, further exploration is needed to extend ML techniques to RF impedance modeling.

## 5 Standardization and Open Benchmark Dataset for Analog/RF ML-EDA

When applying ML to analog design, an important challenge to address is the lack of standardization in data, workflows, and evaluation protocols. For example, analog sizing techniques are commonly evaluated on OTAs that are inconsistently characterized across different technologies. Due to the lack of standardization, a fair comparison between newly proposed techniques is challenging to perform and often erroneously completed [81].

Efforts in digital EDA [81–85] have highlighted the importance of unified frameworks and open datasets for the application of ML to EDA problems. Establishing uniform benchmarks and evaluation criteria for analog design will improve comparability, reproducibility, and model reusability of workflows while addressing limitations due to dataset and PDK disparities.

## 6 Concluding Remarks

ML has proven to be a powerful tool in improving productivity and design quality in analog and RF circuit design, complementing, rather than replacing, traditional techniques. A summary of heuristic and ML-based techniques for each circuit task is shown in Fig. 2. The infusion of ML/AI reduces the traditionally steep learning curve needed in analog and RF design.